\documentclass{article}

\usepackage{arxiv}

\usepackage[utf8]{inputenc} 
\usepackage[T1]{fontenc}    
\usepackage{hyperref}       
\usepackage{url}            
\usepackage{booktabs}       
\usepackage{amsfonts}       
\usepackage{nicefrac}       
\usepackage{microtype}      
\usepackage{graphicx}
\usepackage{amsmath}
\usepackage{amssymb}
\usepackage[capitalise]{cleveref}
\usepackage{latexsym}
\usepackage{algorithm2e}

\title{Exploration of the Applicability of Probabilistic Inference for Learning Control in Underactuated Autonomous Underwater Vehicles}

\author{
  Wilmer Ariza Ramirez\\
  Australian Maritime College\\
  University of Tasmania\\
  Newnham, Australia \\
  \texttt{Wilmer.ArizaRamirez@utas.edu.au} \\
   \And
 Zhi Q. Leong \\
  Australian Maritime College\\
  University of Tasmania\\
  Newnham, Australia \\
  \texttt{Zhi.Leong@utas.edu.au} \\
   \And
 H. Nguyen \\
  Australian Maritime College\\
  University of Tasmania\\
  Newnham, Australia \\
  \texttt{h.d.nguyen@utas.edu.au} \\
     \And
 S.G. Jayasinghe \\
  Australian Maritime College\\
  University of Tasmania\\
  Newnham, Australia \\
  \texttt{shantha.jayasinghe@utas.edu.au} \\
}

\begin{document}
\maketitle

\begin{abstract}
Underwater vehicles are employed in the exploration of dynamic environments where tuning of a specific controller for each task would be time-consuming and unreliable as the controller depends on calculated mathematical coefficients in idealised conditions. For such a case, learning task from experience can be a useful alternative. This paper explores the capability of probabilistic inference learning to control autonomous underwater vehicles that can be used for different tasks without re-programming the controller. Probabilistic inference learning uses a Gaussian process model of the real vehicle to learn the correct policy with a small number of real field experiments. The use of probabilistic reinforced learning looks for a simple implementation of controllers without the burden of coefficients calculation, controller tuning or system identification.  A series of computational simulations were employed to test the applicability of model-based reinforced learning in underwater vehicles. Three simulation scenarios were evaluated: waypoint tracking, depth control and 3D path tracking control. The 3D path tracking is done by coupling together a line-of-sight law with probabilistic inference for learning control. As a comparison study LOS-PILCO algorithm can perform better than a robust LOS-PID. The results shows that probabilistic model based reinforced learning is a possible solution to motion control of underactuated AUVs as can generate capable policies with minimum quantity of episodes.
\end{abstract}

\section*{Acknowledgements}
The authors thank Defence Science and Technology Group for the loan of the vehicle MULLAYA to the Australian Maritime College, and constant support on the platform development.

\section{Introduction}
Autonomous underwater vehicles play an important role in the exploration of the seas. This exploration is primarily driven by commercial, military and scientific needs. In this context, the proper and correct navigation of the vehicle is a key requirement. Motion controllers that are used for navigating AUVs can be classified in four basic strategies: point stabilization \cite{doi:10.5772/61037}, trajectory tracking \cite{980898}, path following \cite{RN1} and path tracking \cite{6094949}. Point stabilization controllers stabilize a vehicle to the desired goal posture from an initial configuration \cite{RN6}. Trajectory tracking controllers use a virtual vehicle to generate a reference trajectory that has an associated time required to be employed by the real vehicle \cite{RN24}. In the case of path following the vehicle is forced to pursue the desired path without temporal specifications. In the case of path following controller, they usually employ the Frenet-Serret line-of-sight(LOS) coupled with another controller to minimize the error between the obtained geometric references and the vehicle variables. The final strategy is path tracking, which combines trajectory tracking and path following by the introduction of a virtual time parameter to force the vehicle to complete the path whitin a specific time.

Difficulty of controlling underwater vehicles arises due to non-linear and time-varying dynamics of underwater vehicles, uncertainties in its hydrodynamic coefficients and disturbance in the environment.(e.g. ocean currents). Furthermore, all complexities are exacerbated for the controller in underactuated vehicles \cite{RN6}; underactuated vehicles have more degrees of freedom to be controlled than surfaces of control. Nevertheless, this configuration is more prevalent as it is the most energy efficient design for travelling at high speeds \cite{XIANG201514}.

Waypoint tracking is the most common methodology to control a vehicle, e.g. commercial vehicles such as Gavia \cite{roper}and REMUS \cite{holsen2015dune} use this methodology. Waypoint tracking is directing the vehicle to approximate to a series of specific target points. The vehicle calculates the required direction to which the vehicle should be directed and upon arriving at the proximities of a point is given a new target. Like many industries, PID is the most common methodology to control the vehicle orientation and speed. However, there is research to employ more robust options than PID.  \cite{7587396} employs a NARMAX model from the vehicle and a constrained self-tuning controller to direct the vehicle to the respective target. Other methodologies employ a combination of LOS for waypoint \cite{ataei2015three} and a standard controller or backstepping techniques to minimize the error between the vehicle position and the desired position  \cite{saravanakumar2011waypoint}.

In the case of LOS, commercially LOS-PID and LOS-Fuzzy controllers are employed as their implementation are simpler and more accepted in the industry \cite{RN29,XIANG2017165,7890302}. Other research as  \cite{RN12} used a LOS guidance law with two integrators and three feedback controllers to compensate for external unknown perturbation such as ocean current which is one of the weaknesses for methodologies as PID/Fuzzy controllers.  \cite{yang2016path} have used an alternative methodology of a grey prediction to obtain the next AUV position in advance and then use LOS to calculate the desired angles, such that if there is environmental interference, the vehicle will not be affected.

In the search of more robust controllers, nonlinear control techniques had been explored. \cite{REPOULIAS20071650} have designed a horizontal path following controller based on Lyapunov stability theorem and backstepping method. In  \cite{liang2015path}, a method consisted of Lyapunov stability theorem and feedback gain backstepping reduce the complexity of the controller and improve adjustability of the parameters. Another methodology proposes a global path following for AUV based on the same coordinates to achieve global asymptotic stability of the following error \cite{gao2010global}. Following the research of backstepping, \cite{doi:10.5772/64065} have adopted fuzzy backstepping sliding mode control to overcome non-linearities, uncertainties and external disturbances.

However, the aforementioned research in controls are focused to provide a more robust path following performance. The controller still requires the calibration of parameters or specific design of observers to identify the unknown parameters of the dynamic model. A methodology to overcome this is the use of machine learning algorithms. The most prominent algorithm in machine learning is neural networks. In particular, the research to control underwater vehicles had focused on the use of machine learning algorithms to recognize uncertainties. \cite{RN28,6196982} have designed a combined version of control law for the convergence of the kinematic model and an adaptive backstepping sliding control based in radial basis function (RBF) neural network to identify the unknown parameters of the dynamic model. In a similar way, \cite{8543571} have reduce the backstepping complexity by the inclusion of a second-order filter to obtain the derivatives of the virtual controller and filter high-frequency measurement noise, and coupled the filter with an RBF neural network that compensates for vehicle uncertainties.

Although the practicality of machine learning has been largely to identify uncertainties in AUV control, some machine learning algorithms are capable of doing more such as controlling the vehicle directly. Recently, there has been increasing research efforts on the use of reinforcement learning to generate policies to control underwater vehicles and robots in general. An example of machine learning control can be seen in \cite{1544973},where reinforced learning (RL) based on the Markov decision process (MDP) was employed to produce a policy capable of controlling a vehicle around an obstacle with a minimum cost. In the case of path-following, reinforced learning had been applied to path following of ships. In  \cite{7483431}, an actor-critic multilayer perception reinforced learning is used to reduce the tracking error to zero. Deep reinforced learning has also been proposed as a possible solution for the tracking problem.  \cite{yu2017deep} employed two neural networks. The primary neural network selects the action and the secondary evaluates whether the produced action is valid; with further modification through a deep deterministic policy gradient. Another application used continuous actor-critic learning automaton algorithm to teach an AUV to follow a pipeline  \cite{fjerdingen2010auv}, considering the improved performance in search of the policy of this algorithm the number of episodes over the platform can be over the hundreds.

RL can be divided into two methodologies: model-based methods and model-free methods, such as Q-learning \cite{watkins1989learning} or TD-learning \cite{sutton2011reinforcement}. The application of path-following control based in traditional RL such as Q-learning  \cite{gaskett1999reinforcement} is highly complex and difficult as a high quantity of experiments is required to acquire data and test each policy iteration. The additional difficulties in underwater vehicles are vehicle safety, maximum time underwater and computational power. RL for a system with low-dimensional state spaces and fairly favourable dynamics can require thousands of trials to arrive at the appropriate policies\cite{6654139,yu2017deep}.

Model-based RL methods are more efficient than model-free methods in searching for a useful policy, as the policy is searched over a model and not the real platform. However, their accuracy can suffer severely from model errors. A solution to address the model errors is the use of probabilistic models to express its uncertainty. An application of model-free methodologies that use a probabilistic methodology was propose by \cite{7401861}. Their methodology use a on-line selective reinforcement learning approach combined with Gaussian Process (GP) regression for learning reference tracking control policies given no prior knowledge of the dynamical system. \cite{6654139} have proposed Probabilistic Inference for Learning Control (PILCO), which is a model-based policy search method. The probabilistic model uses non-parametric Gaussian processes (GPs) to characterise the model uncertainty and the policy improvement is based on analytic policy gradients which employs deterministic approximate interference techniques. Due to probabilistic modelling and inference approach, PILCO can achieve higher learning efficiency than other methods in continuous state-action domains and, hence, is directly applicable to complex mechanical systems, such as robots. 

In this paper, the authors explore the applicability of PILCO to control underactuated AUVs by a series of simulations with different objectives and target values. The main goals of our implementation of reinforced learning with PILCO are:

\begin{itemize}
\item Minimum quantities of episodes over the platform;
\item Small test time over the platform;
\item Minimum quantity of variables to be predicted by the GP; and
\item Vehicle safety.
\end{itemize}

\section{Underwater Vehicle Mathematical Model}
\begin{figure}[!htb]
\centering
\includegraphics[width=3in]{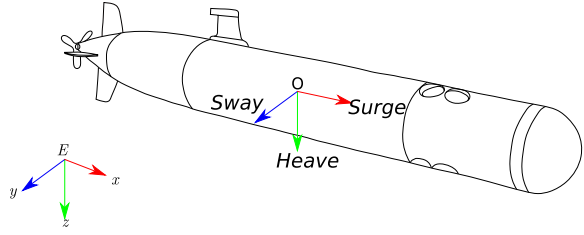}
\caption{AUV different reference frames, vehicle frame is equal to the centre of buoyancy.}
\label{fig:figure-71}
\end{figure}

In \cite{RN76} it was shown that the non-linear dynamic equations of motion of an underwater vehicle can be expressed in vector notation defined by a state vector composed by the vector $v$ of velocities on the body frame of the form ${\left[ {u,v,w,p,q,r} \right]^T}$ and the vector $\eta $ of position in the Earth fixed frame (\cref{fig:figure-71}) of the form  ${\left[ {\xi ,\eta ,\zeta ,\phi ,\theta ,\psi } \right]^T}$ such that
\begin{equation}\label{eq:eq7.1}
{\bf{M\dot v + C}}\left( {\bf{v}} \right){\bf{v + D}}\left( {\bf{v}} \right){\bf{v + g}}\left( {\bf{\eta }} \right){\bf{ = \tau }}
\end{equation}
with the kinematic equation
\begin{equation}
\label{eq7.2}
{\bf{\dot \eta  = J}}\left( {\bf{\eta }} \right){\bf{v}}
\end{equation}
where

${\bf{\eta }}$ \ \ position and orientation of the vehicle in Earth-fixed frame,

${\bf{v}}$ \ \ linear and angular vehicle velocity in body fixed frame,

${\bf{\dot v}}$ \ \ linear and angular vehicle acceleration in body fixed frame,

${\bf{M}}$ \ \ matrix of inertial terms,

${\bf{C}}\left( {\bf{v}} \right)$ \ \ matrix of Coriolis and centripetal terms,

${\bf{D}}\left( {\bf{v}} \right)$ \ \ matrix consisting of damping or drag terms,

${\bf{g}}\left( {\bf{\eta }} \right)$ \ \ vector of restoring forces and moments due to gravity and buoyancy,

${\bf{\tau }}$ \ \ vector of control and external forces, and

${\bf{J}}\left( {\bf{\eta }} \right)$ \ \ rotation matrix that converts velocity in a body fixed frame  $v$ \ to an Earth fixed frame velocity  $\dot \eta $ .

\
\Cref{eq:eq7.1} can be expanded into a more general equation of motion as has been shown in \cite{RN178,gertler1967standard}. The result of the expansion will be a system of six equation with 73 hydrodynamic coefficients. However, for a complete model the control surfaces must be modelled. In a general case, the resulting forces and moments of a control surface (thrusters and fins) can be expressed as \cite{RN76}
\begin{equation}\label{eq7.3}
\begin{array}{l}
{F_{prop}} =  - {K_{fprop}}\left| n \right|n\\
{M_{prop}} =  - {K_{mprop}}\left| n \right|n
\end{array}
\end{equation}
\begin{equation}\label{eq7.4}
\begin{array}{l}
{L_{fin}} = {K_{\left. L \right|{\delta _{{\rm{fin}}}}}}{\delta _{{\rm{fin}}}}v_e^2\\
{M_{fin}} = {K_{\left. M \right|{\delta _{{\rm{fin}}}}}}{\delta _{{\rm{fin}}}}v_e^2
\end{array}
\end{equation}

A more accurate thruster model can be found \ in \cite{kim2006accurate} with the inclusion of the motor model and fluid dynamics. However, in this study, the more conservative model from \cite{RN76} is used.

\section{LOS guidance law mathematical background}\label{LOS}
This section describes the mathematical background of the 3D guidance law employ in the present study for control of an underactuated AUV. In this study, it was decided to employ the LOS proposed in  \cite{XIANG2017165} as it is an extension of the work of \cite{RN76}.

\begin{figure}[htbp]
\centering
\includegraphics[width=3in]{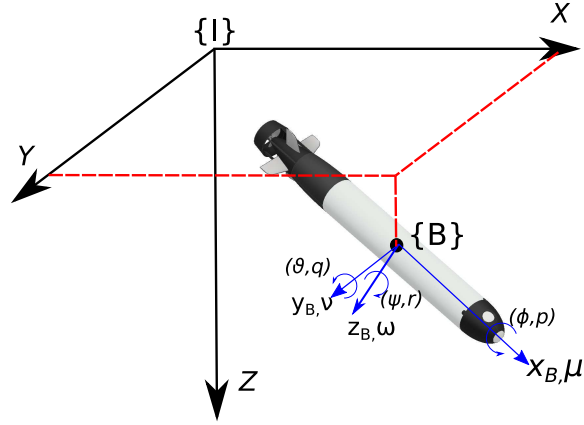}
\caption{LOS reference frames}
\centering
\label{fig:p7fig2}
\end{figure}

\begin{figure}[htbp]
\centering
\includegraphics[width=3in]{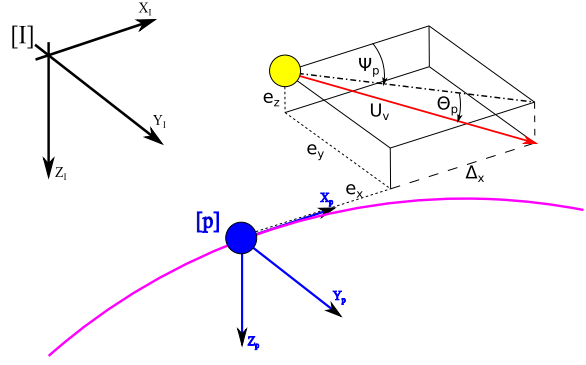}
\caption{LOS main variables}
\centering
\label{LOSpoints}
\end{figure}

If the vehicle kinematic is represented by its spatial position $p\left( t \right) \buildrel \Delta \over = {\left[ {x\left( t \right),y\left( t \right),z\left( t \right)} \right]^T}$ and its velocity is represented  by $v(t) \triangleq \dot p\left( t \right) \in {\mathbb{R}^3}$ state is related to the $\left\{ I \right\}$ frame. Also, the speed is represented  by $U\left( t \right) \buildrel \Delta \over = \left| {v\left( t \right)} \right| = \sqrt {\dot x{{\left( t \right)}^2} + \dot y{{\left( t \right)}^2} + \dot z{{\left( t \right)}^2}}  > 0$. The steering of the underactuated AUV is characterized by the azimuth angle $\chi$ and elevation $\upsilon$ (\cref{fig:p7fig2}). 
\begin{equation}\label{eq7.5}
\left\{ {\begin{array}{*{20}{c}}
{\chi  = {\rm{atan2}}\left( {\dot y,\dot x} \right)}\\
{\upsilon  = \arctan \left( {\frac{{ - \dot z}}{{\sqrt {{{\dot x}^2} + {{\dot y}^2}} }}} \right)}
\end{array}} \right.
\end{equation}
If its consider a continuously path parametrized by a scalar variable $\varpi \in {\mathbb{R}}$, the position of a point over the path is represented by ${p_p}\left( \varpi  \right) \in {\mathbb{R}^3}$ (\cref{LOSpoints}). Similarly, the orientation of the point can be defined as
\begin{equation}\label{eq7.6}
\left\{ {\begin{array}{*{20}{c}}
{{\psi _p} = {\rm{atan2}}\left( {{{y'}_p},{{x'}_p}} \right)}\\
{{\theta _p} = \arctan \left( {\frac{{ - {{z'}_p}}}{{\sqrt {{{x'}^2}_p + {{y'}^2}_p} }}} \right)}
\end{array}} \right.
\end{equation}
where ${x'}_p=dx_p/d\varpi$, ${y'}_p=dy_p/d\varpi$ and ${z'}_p=dz_p/d\varpi$ . Hence the tracking error is expressed as:

\begin{equation}\label{eq7.7}
\varepsilon  = {\left[ {{x_e},{y_e},{z_e}} \right]^{\rm T}} = {\mathbf{R}}_F^{\rm T}\left( {p - {p_p}} \right)
\end{equation}

where ${\mathbf{R}}_F^{\rm T}: = {{\mathbf{R}}_z}\left( {{\psi _p}} \right){{\mathbf{R}}_y}\left( {{\theta _p}} \right)$

\begin{equation}\label{eq7.8}
{{\mathbf{R}}_z} = \left[ {\begin{array}{*{20}{c}}
  {\cos \left( {{\psi _p}} \right)}&{ - \sin \left( {{\psi _p}} \right)}&0 \\ 
  {\sin \left( {{\psi _p}} \right)}&{\cos \left( {{\psi _p}} \right)}&0 \\ 
  0&0&1 
\end{array}} \right]
\end{equation}

\begin{equation}\label{eq7.9}
{{\mathbf{R}}_z} = \left[ {\begin{array}{*{20}{c}}
  {\cos \left( {{\theta _p}} \right)}&0&{\sin \left( {{\theta _p}} \right)} \\ 
  0&1&0 \\ 
  { - \sin \left( {{\theta _p}} \right)}&0&{\cos \left( {{\theta _p}} \right)} 
\end{array}} \right]
\end{equation}

If the following Lyapunov control function positive definitive is 
\begin{equation}\label{eq7.10}
{V_\varepsilon } = \frac{1}{2}{\varepsilon ^T}\varepsilon 
\end{equation}
The derivative of \cref{eq7.10} can be written as
\begin{equation}\label{eq7.11}
\begin{gathered}
  {V_\varepsilon } = {x_e}\left( {{U_d}\cos \left( {{\psi _r}} \right)\cos \left( {{\theta _r}} \right) - {U_p}} \right) +  \\ 
  {y_e}{U_d}\sin \left( {{\psi _r}} \right)\cos \left( {{\theta _r}} \right) - {z_e}{U_d}\sin \left( {{\theta _r}} \right) \\ 
\end{gathered}
\end{equation}
where $U_d$ is the desired composite speed of the AUV. The auxiliary control input of the virtual point $P_p$ is chosen as:
\begin{equation}\label{eq7.12}
U_p={U_d}\cos(\psi_r)\cos(\theta_r)+{k_x}{x_e}
\end{equation}
where the steering angles are:
\begin{equation}\label{eq7.13}
\left\{ {\begin{array}{*{20}{c}}
{{\psi _r} = \arctan \left( {\frac{{ - {k_y}{y_e}}}{{{\Delta _y}}}} \right)}\\
{{\theta _r} = \arctan \left( {\frac{{{k_z}{z_e}}}{{{\Delta _z}}}} \right)}
\end{array}} \right.
\end{equation}
If the guidance variables $\Delta_y$ and $\Delta_z >0$, the control gains ${k_x},{k_y},{k_z}$ are positive constants. If \cref{eq7.12} and \cref{eq7.13} are substituted into \cref{eq7.11} and considering the relationship among inertial frame $\left\{ I \right\}$, flow frame $\left\{ W \right\}$ and path frame $\left\{ F \right\}$, the desired azimuth angle $\upsilon_d$ and elevation angle $\chi_d$ can be written as \cite{Fossen09}:
\begin{equation}\label{eq7.14}
\begin{array}{l}
{\upsilon _d} = \arcsin (\sin {\theta _p}\cos {\psi _r}\cos {\theta _r} + \cos {\theta _p}\cos {\theta _r})\\
{\chi _d} = atan2({\chi _d}_y,{\chi _d}_x)
\end{array}
\end{equation}
where
\begin{equation}
\begin{array}{c}
{\chi _d}_y = \cos {\psi _p}\sin {\psi _r}\cos {\theta _r} - \sin {\psi _p}\sin {\theta _p}\sin {\theta _r}\\
 + \sin {\psi _p}\cos {\theta _p}\cos {\psi _p}\cos {\theta _r}
\end{array}
\end{equation}

\begin{equation}
\begin{array}{c}
{\chi _d}_x = -\sin {\psi _p}\sin {\psi _r}\cos {\theta _r} - \cos {\psi _p}\sin {\theta _p}\sin {\theta _r}\\
 + \cos {\psi _p}\cos {\theta _p}\cos {\psi _p}\cos {\theta _r}
\end{array}
\end{equation}
In order to transform the path following to path tracking the path was defined over time together with \cref{eq7.14} and \cref{eq7.12}. This  was to produce not only the desire angles but also the required speed at each time instance. The steering error vector can be expressed as $[e_\mu, e_\upsilon,e_\chi]$ where $e_\mu=\mu_d-\mu_v$, $e_\upsilon=\upsilon_d-\upsilon_v$ and $e_\chi=\chi_d-\chi_v$.
\section{Probabilistic Inference for Learning Control (PILCO)}
PILCO algorithm \cite{6654139,deisenroth2011pilco} (\cref{PILCO}) employs GPs as the base for policy search. A GP can be defined by a mean function $m(\cdot)$ and a positive definitive covariance function $k(\cdot,\cdot)$ commonly known as kernel. Usually a prior mean function $m\equiv0$ and an exponentiated quadratic kernel (\cref{eq7.15}) are employ. This kernel only has two parameters to learn, $l$ that determines the length of the 'wiggles' in the function and $\sigma^2$ which determines the average distance of the function away from its mean \cite{duvenaud2014automatic}.
\begin{equation}\label{eq7.15}
k_{\textrm{SE}}(x, x') = \sigma^2\exp\left(-\frac{(x - x')^2}{2\ell^2}\right)
\end{equation}
Given $n$ training inputs $X=\left[x_1,...,x_n\right]$ and corresponding training targets $Y=\left[y_1,...,y_n\right]$, the posterior GP hyper-parameters $l$ and $\sigma^2$ are learned by evidence maximization \cite{rasmussen2004gaussian}.
The posterior predictive distribution $p(f_*|x_*)$ of the function value $f_*=f(x_*)$ for a test input $x_*$ is Gaussian with mean and variance
\begin{equation}
\begin{gathered}
  {m_f}\left( {{x_*}} \right) = {\mathbf{k}}_*^T{\left( {{\mathbf{K}} + \sigma _\varepsilon ^2{\mathbf{I}}} \right)^{ - 1}}y \hfill \\
  \sigma _f^2\left( {{x_*}} \right) = {k_{**}} - {\mathbf{k}}_*^T{\left( {{\mathbf{K}} + \sigma _\varepsilon ^2{\mathbf{I}}} \right)^{ - 1}}{k_*} + \sigma _\varepsilon ^2 \hfill \\ 
\end{gathered} 
\end{equation}

\begin{algorithm}[htbp]
\SetNlSty{texttt}{[}{]}
\SetAlgoNlRelativeSize{0}
\SetNlSkip{0em}
\nl\emph{\textbf{init:}Sample controller parameters $\theta \sim \mathcal{N}(0,\,I)$}\;
\nl\emph{Apply random control signals and record data.}\;
\nl\Repeat{Task Learned}{
\nl\emph{Learn probabilistic (GP) dynamics model}\;
\nl\Repeat{Convergence}{
\nl\emph{Approximate inference for policy evaluation}\;
\nl\emph{Gradient-based policy improvement}\;
\nl\emph{Update parameters $\theta$}\;
}
\nl\KwRet{${\theta ^ * }$}\;
\nl\emph{Set ${\pi ^ * } \leftarrow \pi \left( {{\theta ^ * }} \right)$}\;
\nl\emph{Apply $\pi ^ *$ to system and record data}\;
}

\caption{PILCO Algorithm}\label{PILCO}
\end{algorithm}
\DecMargin{1em}

To find a policy $\pi^*$ which converges to the desired target, PILCO builds a probabilistic GP dynamics model. This model will be the base for the deterministic approximate inference and policy evaluation, followed by the analytic computation of the policy gradients $\partial{J^\pi(\theta)}/\partial{\theta}$ for policy improvement. The policy $\pi$ is improved based on the gradient information $\partial{J^\pi(\theta)}/\partial{\theta}$.

\section{Simulation Setup}
The underwater vehicle should be able to switch between different controllers in a single mission depending on the task to be solved at a specific time. The possible variants of controllers that an AUV can use in a mission are: follow bottom, depth control, follow pipe, go-to-point, path tracking, path following, etc. A series of different simulations were designed to evaluate the capability of PILCO to control an underactuated AUV within the three different evaluation scenarios: way-point-tracking, Depth Control and Path-Tracking.

\subsection{Vehicle Model}
The vehicle model employed in this research is derived from semi-empirical calculation of the coefficients for the vehicle MULLAYA. MULLAYA is an underactuated AUV designed by the Defence Science and Technology Group as a research platform. The vehicle is controlled by a single propeller, a pair of elevator fins and a pair of rudder fins. General specifications of the vehicle are given in \cref{MULLAYATABLE}. The coefficients were calculated with the same technique of \cite{RN178} and the obtained coefficients are presented in \cref{tab:MullayaCoeficient}. As a engineering research platform, the vehicle will undergo multiple transformations over time in shape and internal engineering. This constant change requires the needs of constant update of the controller for the vehicle. Possible future modification can include vectorized propulsion systems and buoyancy controllers.

\begin{table}[htbp]
  \centering
  \caption{MULLAYA AUV particulars.}
    \begin{tabular}{c|c|c}
    \textbf{Property} & \textbf{Value} & \textbf{Unit} \\
    \midrule
    Length & 1.56  & \textbf{m} \\
    Diameter & 150   & \textbf{mm} \\
    Max RPM & 3000  & \textbf{RPM} \\
    Weight & 239.364 & \textbf{N} \\
    Buoyancy & 246.2 & \textbf{N} \\
    \end{tabular}%
  \label{MULLAYATABLE}%
\end{table}%

\begin{table}[htbp]
  \centering
  \caption{MULLAYA AUV coefficients employed in simulations.}
    \begin{tabular}{c|c|c|c|c|c}
    \textbf{Coeff.} & \textbf{Result} & \textbf{Coeff.} & \textbf{Result} & \textbf{Coeff.} & \textbf{Result} \\
    \hline
    Xuu   & -2.8 & Zrp   & 0.68 & Nuv   & -30.88 \\
    Xwq   & -29.6 & Yuudr & 12.12 & Npq   & -5.06 \\
    Xqq   & -0.68& Nuudr & -7.51 & Ixx   & 0.083 \\
    Xvr   & 29.6 & Zuuds & -12.12 & Iyy   & 3.08 \\
    Xrr   & -4.95 & Kpp   & -1.30E-01 & Izz   & 3.08 \\
    Yvv   & -95.37 & Kpdot & 1.09E-02 & Nwp   & 0.66 \\
    Yrr   & -2.45 & Mww   & 6.96 & Nur   & -5.31 \\
    Yuv   & -32.9 & Mqq   & -135.04 & Xudot & -0.51 \\
    Ywp   & 29.64 & Mrp   & 5.1 & Yvdot & -29.64 \\
    Yur   & 7     & Muq   & -5.34 & Nvdot & 0.66 \\
    Ypq   & 0.68 & Muw   & 27.16 & Mwdot & 0.66 \\
    Zww   & -95.37 & Mwdot & -0.68 & Mqdot & -5.05 \\
    Zqq   & 2.45 & Mvp   & -0.68 & Zqdot & -4.94 \\
    Zuw   & -32.9 & Muuds & -7.738 & Zwdot & -29.64 \\
    Zuq   & -7    & Nvv   & 6.96 & Yrdot & 4.95 \\
    Zvp   & -29.64 & Nrr   & -135.03 & Nrdot & 5.05 \\
    \end{tabular}%
  \label{tab:MullayaCoeficient}%
\end{table}%

\subsection{Waypoint Tracking}
The first type of controller evaluated for the underwater vehicle is a waypoint tracking controller. These types of controllers can be employed to transfer the vehicles between location were more specific controllers are employed or if can be work with the results of a path planning law that convert a desire path in waypoints. In this case, it is desired to control the vehicle velocity, azimuth and elevation such that the vehicle moves towards a specific location. The methodology used in this simulation is similar to the one employed in  \cite{doi:10.1080/01691864.2014.888373} but extended to 3D. If $\mathbf{\eta_v} $ is the vehicle state vector $[X_v,Y_v,Z_v,\phi_v,\theta_v,\psi_v]$ that can be divided in position vector $\mathbf{X_v}$ and orientation vector $\mathbf{\theta_v}$ on earth frame and the target point is expressed as the vector $\mathbf{X_T}=[X_d,Y_d,Z_d]$ the angles of the vector between the current position and the desired position can be expressed as
\begin{equation}
\begin{gathered}
  {\psi _d} = {\tan ^{ - 1}}\left( {\frac{{{Y_d} - {Y_v}}}{{{X_d} - {X_v}}}} \right) \hfill \\
  {\theta _d} = {\tan ^{ - 1}}\left( {\frac{{\sqrt {{{\left( {{Y_d} - {Y_v}} \right)}^2} + {{\left( {{X_d} - {X_v}} \right)}^2}} }}{{{Z_d} - {Z_v}}}} \right) \hfill \\ 
\end{gathered} 
\end{equation}
and the vector of angle errors can be expressed as the difference between the vehicle orientation and the desire orientation, i.e $\mathbf{\theta_e}=[e_\psi,e_\theta,e_u]$, where $e_\psi=\psi _d-\psi _v$ and $e_\theta=\theta _d-\theta _v$. The target of the policy will be to minimize the error angles to zero and the surge speed error to zero. In the proposed simulation a surge speed of 1.2$m/s$ was set as the desired speed with an initial position of $[0,0,0]$ ,an orientation of $[0,0,\pi/4]$ and an initial surge speed of $0.5m/s$. The first 12 seconds of real model simulation was employed to learn the policy with a target point with coordinates $[40,40,10]$. A constraint to limit the vehicle turn $<180^o$ was applied to the process of policy testing over the vehicle model. A second simulation with two target points $[30,30,10]$ and $[90,100,10]$ is employed to check the viability of the policy. A total of 1000 sparse points was located as the limiter from which a sparse model will be employ.  The policy and simulation were executed at 5$Hz$. A noise with variance of $\sigma^2=0.005$ was employed in the simulation and a variance of $\sigma^2=0.2$ was employed for the start position of the vehicle. When the vehicle arrives at 3 meters of the objective the vehicle will be given the secondary target as the new objective. If the vehicle arrives at 1 meter of the final target the simulation will stop.
\begin{figure*}[htbp]
\centering
\includegraphics[width=0.8\textwidth]{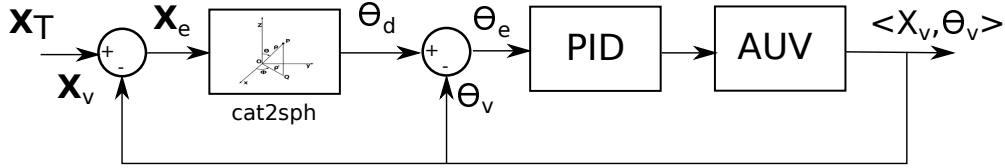}
\caption{Control system block for waypoint tracking with PID controller.}
\centering
\label{waypointpid}
\end{figure*}
\begin{figure*}[htbp]
\centering
\includegraphics[width=0.8\textwidth]{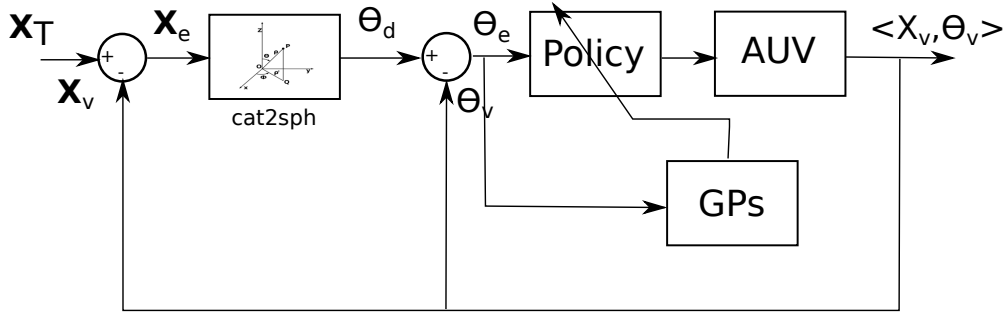}
\caption{Control system blocks for waypoint tracking with RL policy.}
\centering
\label{waypointpilco}
\end{figure*}

\subsection{Depth Control}
A secondary scenario is explored to evaluate the controller’s capability to keep a specific depth while traveling between location. This is an extension to the previous search of a policy, i.e. waypoint. In this scenario, we want the vehicle to arrive at a specific depth before going to the desired location. The policy will minimize the vector of error to zero with the difference that the vector of errors is $\mathbf{\theta_e}=[e_\psi,e_\theta,e_u,e_z]$, where $e_z=z _d-z_v$ similar to the waypoint tracking. The vehicle starts from the same initial position of the waypoint tracking scenario $[0,0,0]$. the proposed training target was the vector  $[40,40,5]$ and the selected test target points are $[30,30,5]$ and $[90,100,5]$. A total of 1500 sparse points was located as the limiter from which a sparse model will be employed. The policy and simulation were executed at 5$Hz$. A noise of $\sigma^2=0.005$ was employed for the simulation and a $\sigma^2=0.2$ was employed for the start position of the vehicle. When the vehicle arrives at 3 meters of the objective the vehicle will be given the secondary target as the new objective. I If the vehicle arrives at 1 meter of the final target the simulation will stop.

\subsection{Path Tracking}
\begin{figure*}[htbp]
\centering
\includegraphics[width=0.8\textwidth]{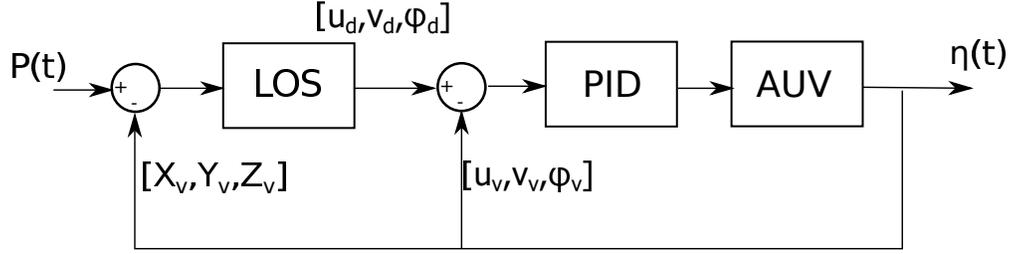}
\caption{Control systems block for LOS-PID}
\centering
\label{LOSPID}
\end{figure*}

\begin{figure*}[htbp]
\centering
\includegraphics[width=0.8\textwidth]{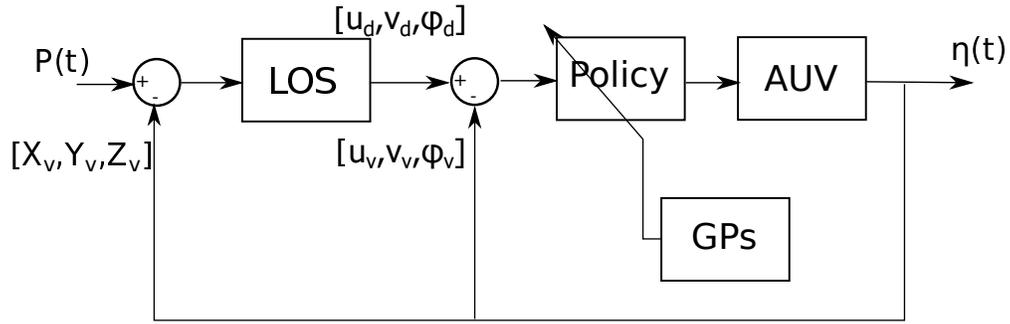}
\caption{Control systems block for LOS-PILCO}
\centering
\label{LOSPILCO}
\end{figure*}

\begin{figure*}
\centering
\includegraphics[width=0.8\textwidth]{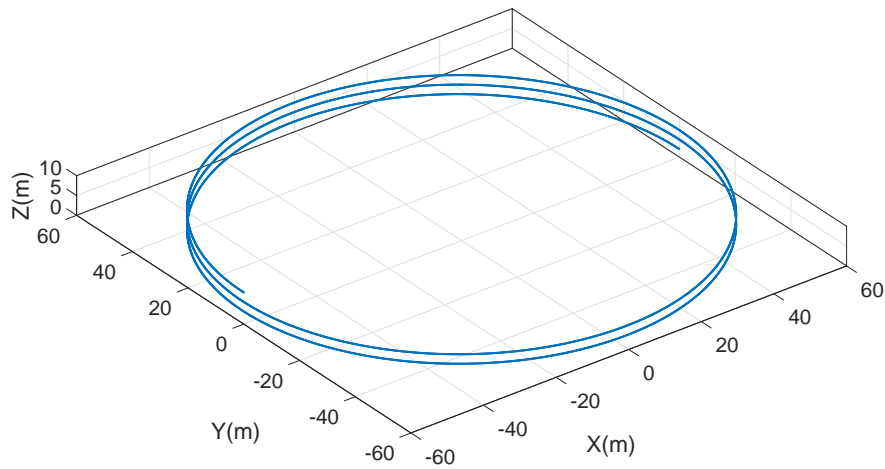}
\caption{Spiral Path to be follow}
\centering
\label{path}
\end{figure*}

With the aim to test the capability of PILCO for path tracking, the scenario consists of a single policy to control propeller force, elevator force and rudder force of the vehicle. The decision to learn the force and not to control direct the RPM and angle of the fins was to be able to compare the policy to a standard controller from the literature. In the design of the authors path tracking simulation, the equation presented in section \cref{LOS}. \cref{LOSPILCO} shows the block diagram of the LOS-PILCO control implementation whereby the policy will evolve based on the learned GPs. The target of all learned policies is to reduce the vector of errors $[e_\mu, e_\upsilon,e_\chi]$ to zero. A LOS-PID was used as a performance comparison. The LOS-PID    (\cref{LOSPID})with the exact coefficient of the model was coded in the same ways as \cite{XIANG2017165} with the inclusion of measurement noise. The initial position of the vehicle was established as $[60,3,1]$ with an initial orientation  $[0,0,3\pi/4]$. A total of 1500 sparse points was located as the limiter from which a sparse model will be employed. A noise with $\sigma^2=0.001$ was employed throughout the simulation and a random variation of the start point with a variance of $\sigma^2=0.2$ was employed. Higher values of $\sigma$ were not possible for the LOS-PID to be comparable as it was not able to overcome higher values of noise. Both the PID and PILCO policy were executed at a frequency of 10 Hz. A helix path (\cref{path}) was parametrized as is show in \cref{spiral}.

\begin{equation}
\begin{gathered}
  {X_{helix}} = 60*\cos (0.02618*{w_{ramp}(t)}) \hfill \\
  {Y_{helix}} = 60*\sin (0.02618*{w_{ramp}(t)}) \hfill \\
  {Z_{helix}} = 2 + 2*{w_{ramp}(t)}/200 \hfill \\ 
\end{gathered}
\label{spiral}
\end{equation}
where $w_{ramp}(t)$ is a function over time with a slope $m$. The control learning of the vehicle was done over the first 20 seconds at a frequency of 10 Hz.

\section{Results}

\subsection{Waypoint Tracking}
\begin{figure*}[htbp]
\centering
\includegraphics[width=0.8\textwidth]{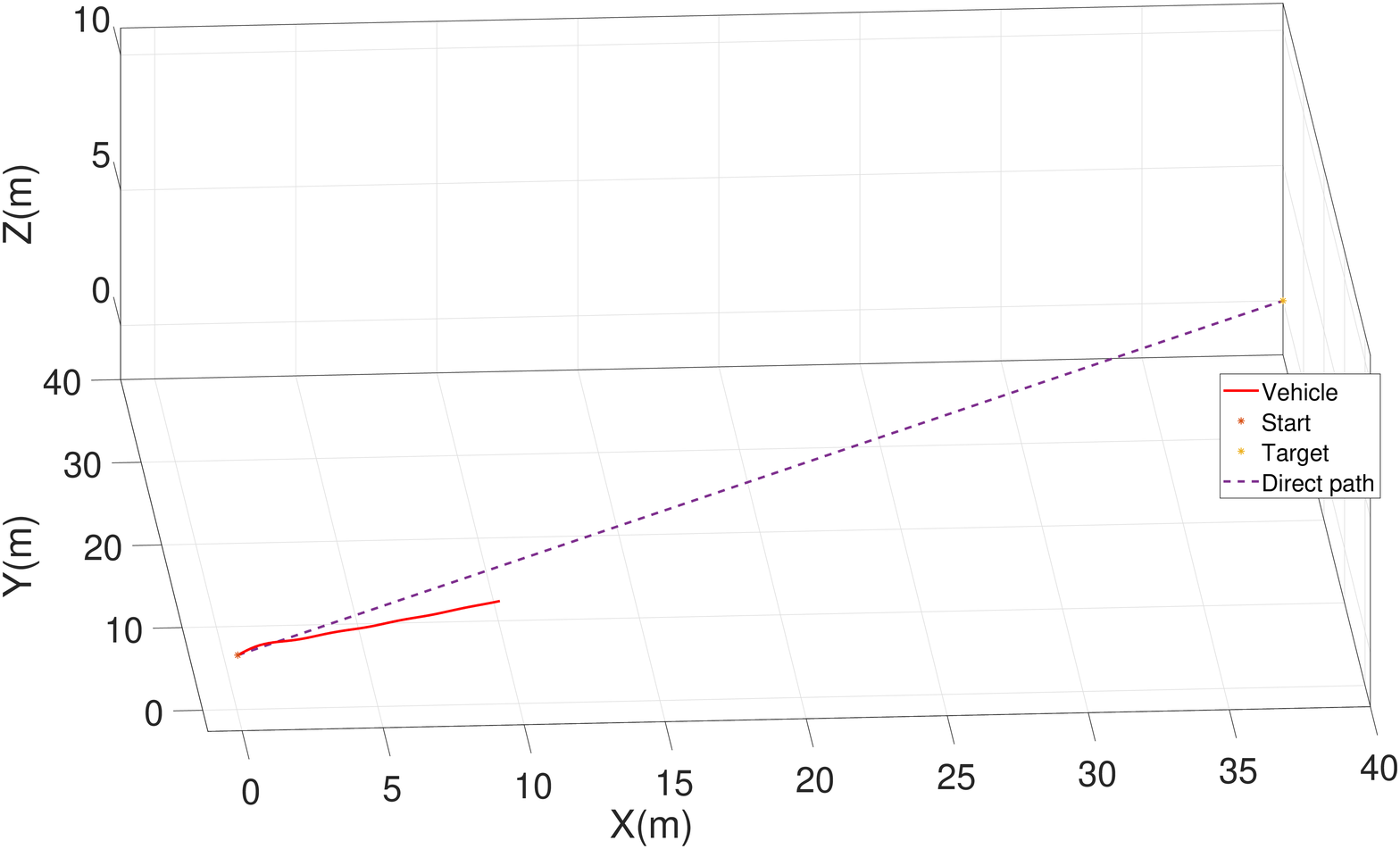}
\centering\caption{Real model waypoint tracking data example for GPs learning.}
\centering
\label{Waypointracking}
\end{figure*}

\begin{figure*}[htbp]
\centering
\includegraphics[width=0.8\textwidth]{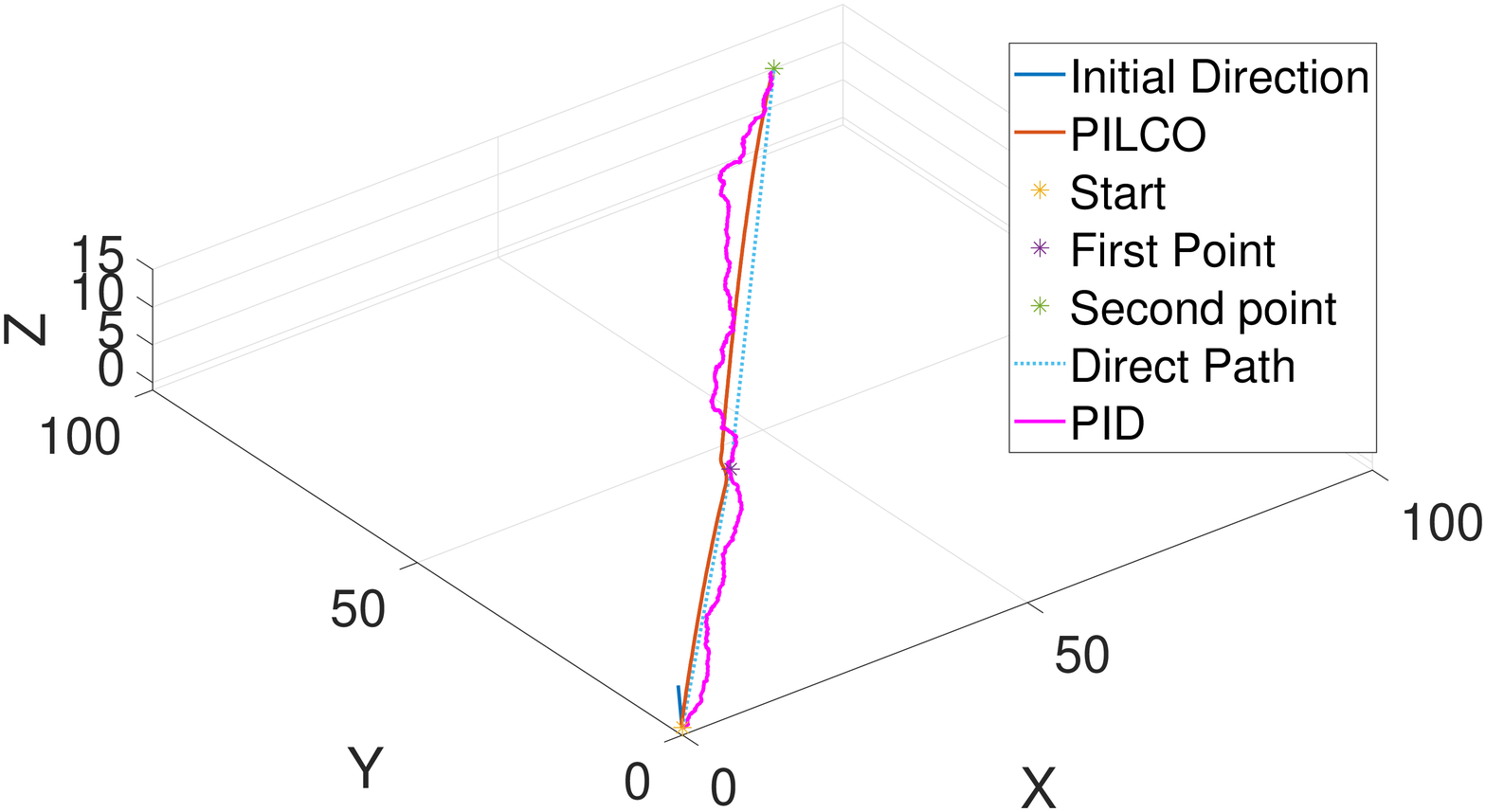}
\centering\caption{3D view Waypoint tracking comparative result between PILCO and PID Results [$m$].}
\centering
\label{WaypointrackingR}
\end{figure*}

As evident in \cref{WaypointrackingR}, the learning of a policy for waypoint tracking of an AUV is possible with the application of PILCO. However, reinforced learning does not require the calibration of parameters, but rather, requires the tuning of the parameters of the $\mathbf{Q}$ matrix from the cost function. For the cost function design, the surge speed was given a higher importance in the cost function than the other error vectors. This escalation of each target in the learning policy was needed as an underactuated AUV needs to get to a specific speed to be able to dive and navigate with a more linear model. An example of the data employed to create the GPs model can be seen in \cref{Waypointracking}.

\begin{figure*}[htbp]
\centering
\includegraphics[width=0.8\textwidth]{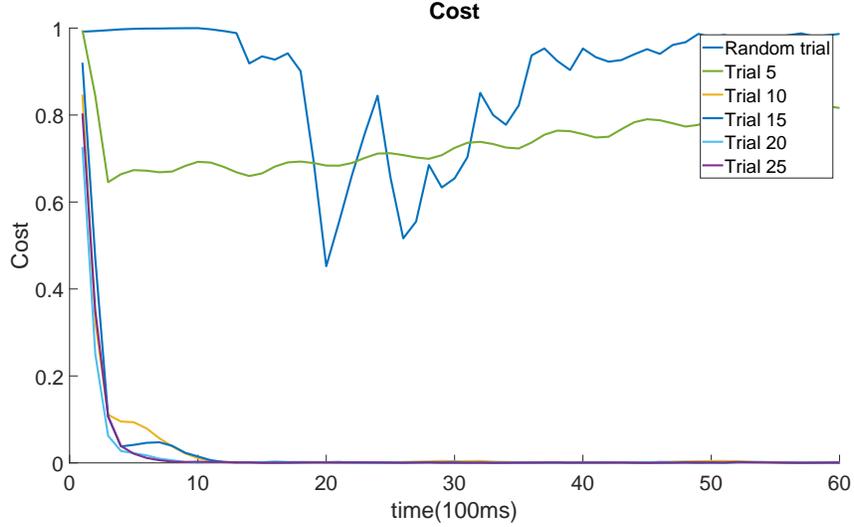}
\centering\caption{Waypoint tracking cost function evolution.}
\centering
\label{WaypointrackingEpoch}
\end{figure*}

The authors simulations showed that the required number of episodes over the platform was under 25 to obtain a usable policy to control the vehicle and be able to arrive at all three targets. \cref{WaypointrackingEpoch} shows the evolution of the cost function over time for the platform. The cost function drops rapidly to a low value as the vehicle learns to control the surge speed and from there, how to control its orientation. The decision to do this is based in that the control surfaces as fin require a minimum speed to be useful.

The result from the final policy selected is shown in \cref{WaypointrackingR}; plotting against the results from the PID controller for the same scenario. Both PID and PILCO arrive near to both test point. However, PILCO shows a more direct route taken towards the targets.The PID controller took 1201 cycles to arrive equivalent to 240.2 seconds and the PILCO policy took 569 cycles equivalent to 113.8 seconds. PILCO has an advantage over a simple PID as PILCO has learned the policy over a noisy platform(or enviroment) and the PID doesn't have any component to compensate for the noise in the measurement. For PID to compensates for the noise an observer will have to be designed and implemented. However, this will increase the complexity of the controller and the need for more design and calibration time for the PID controller. The results show that the waypoint tracking with a PILCO controller can be a viable option as the tuning of the controller coefficients are practically zero. The robustness of the PILCO policy is higher as the platform will start the learning from different angles and can overcome noise in the measurement. The policy is also updated in case of failure. A down point of PILCO in  the simulated scenarios is that the controller tries to always have the same behaviour, e.g. if the vehicle dives a little before directing to the first point after the second point the policy will try to repeat the same behaviour. 

\subsection{Depth control}
\begin{figure*}[htbp]
\centering
\includegraphics[width=0.8\textwidth]{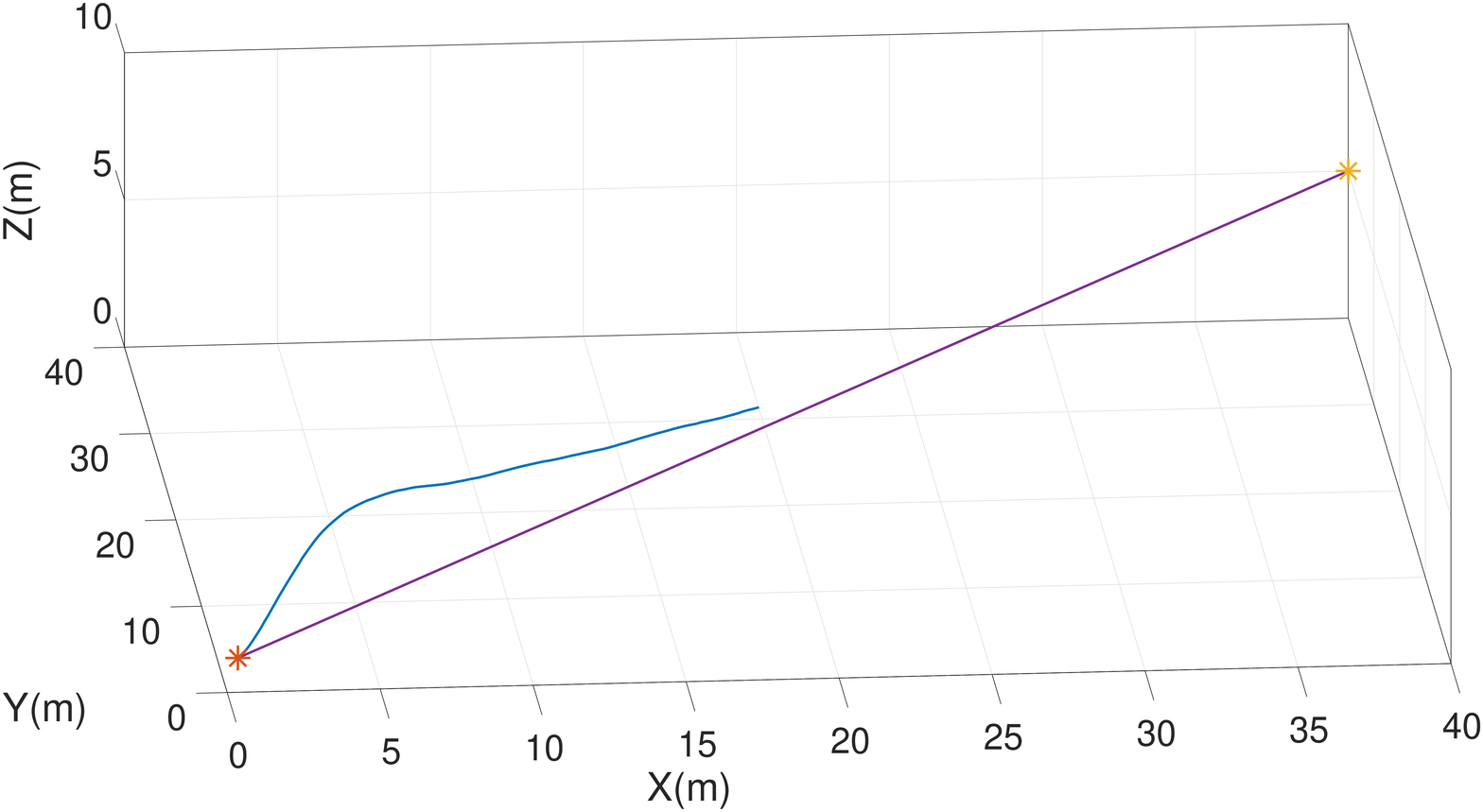}
\centering\caption{Real model waypoint tracking and depth control data example for GPs learning.}
\centering
\label{SAMPLEDepthControl3D}
\end{figure*}

\begin{figure*}[htbp]
\centering
\includegraphics[width=0.8\textwidth]{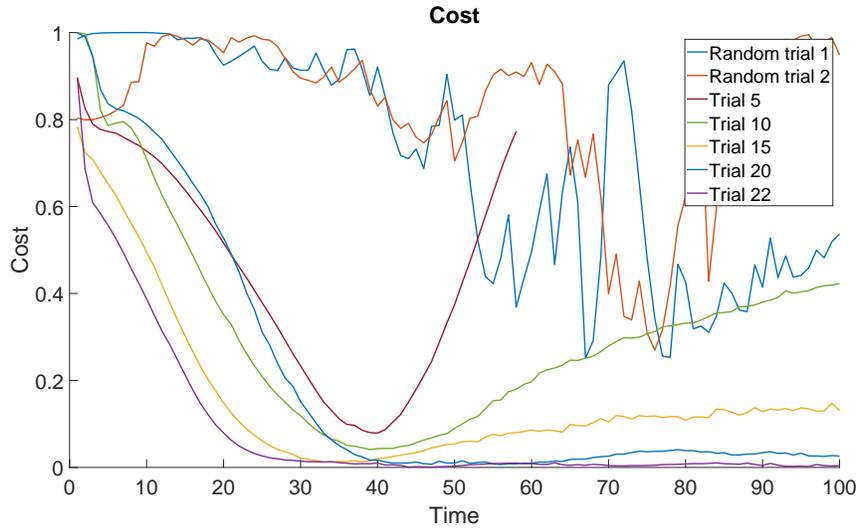}
\centering\caption{Waypoint tracking and depth control cost function evolution.}
\centering
\label{DepthControlepochs}
\end{figure*}

The simulation of simultaneous waypoint tracking, and depth control of underwater vehicles shows that PILCO can learn a complex policy to control the vehicle. An example of the training data used for learning the vehicle GPs model can be seen in \cref{SAMPLEDepthControl3D}. The quality of the learning of the GPs model is directly related to the quality of the learning of the policy. After 20 episodes episodes over the platform a good quality GPs model has been learned such that a policy can be learnt. The evolution of the cost function over time can be seen in \cref{DepthControlepochs}. In similar way to the standard waypoint tracking, the cost evolves very fast to a low value and then small changes are done  in the policy learning. This small change is produced by the scaling of each of the targets to be obtained.

\begin{figure*}[htbp]
\centering
\includegraphics[width=0.8\textwidth]{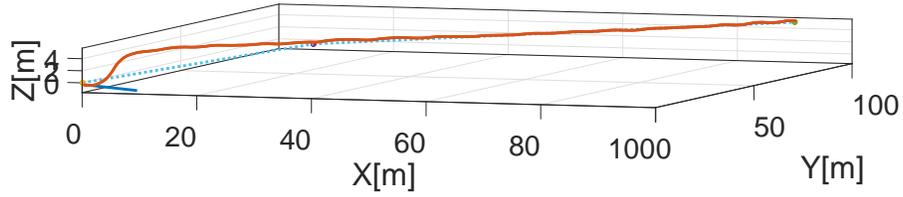}
\centering\caption{3D view Waypoint tracking and depth control PILCO results.}
\centering
\label{DepthControl3D}
\end{figure*}

\begin{figure*}[htbp]
\centering
\includegraphics[width=0.8\textwidth]{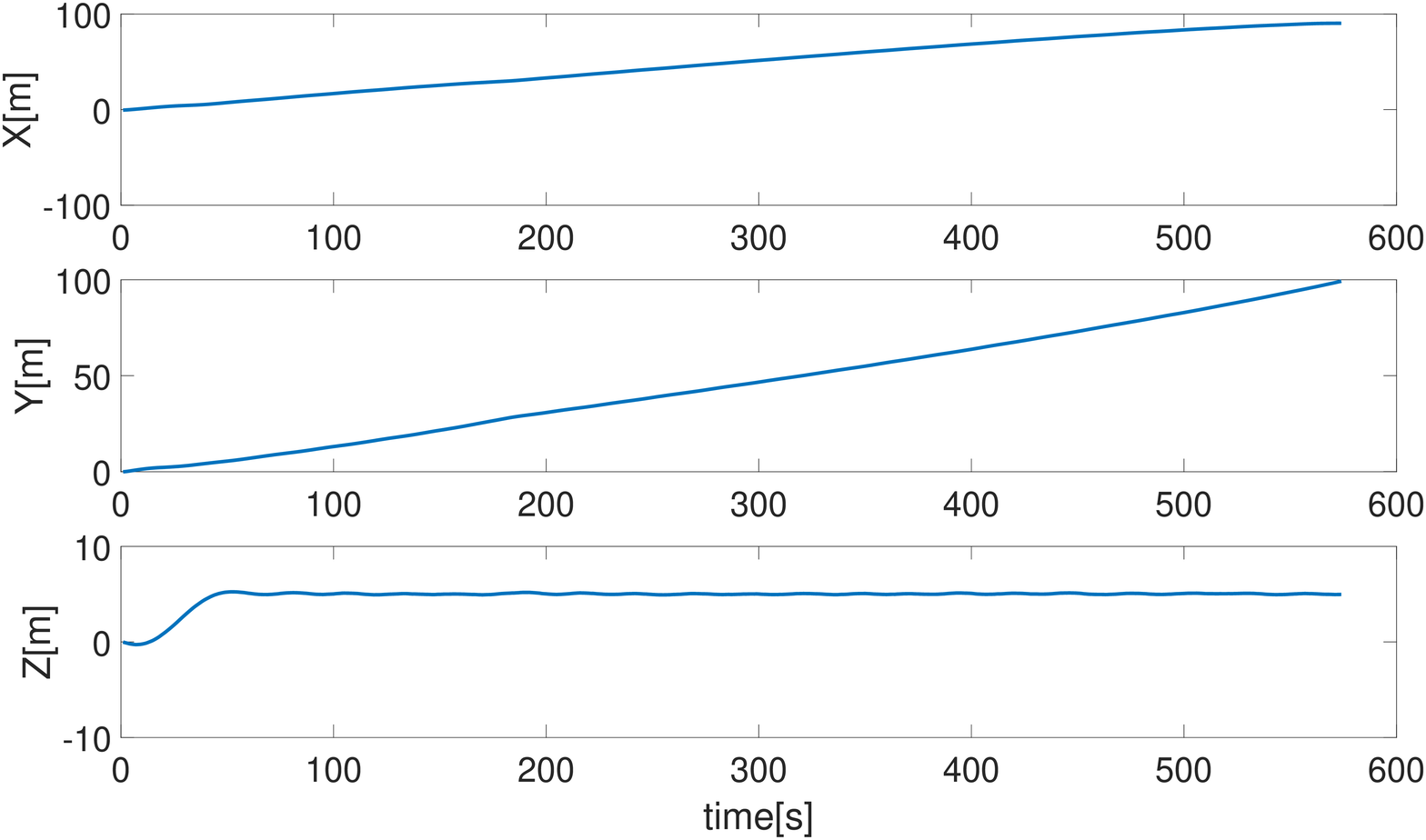}
\centering\caption{$[X,Y,Z]$ Waypoint tracking and depth control PILCO results.}
\centering
\label{DepthControlXYZ}
\end{figure*}

\begin{figure*}[htbp]
\centering
\includegraphics[width=0.8\textwidth]{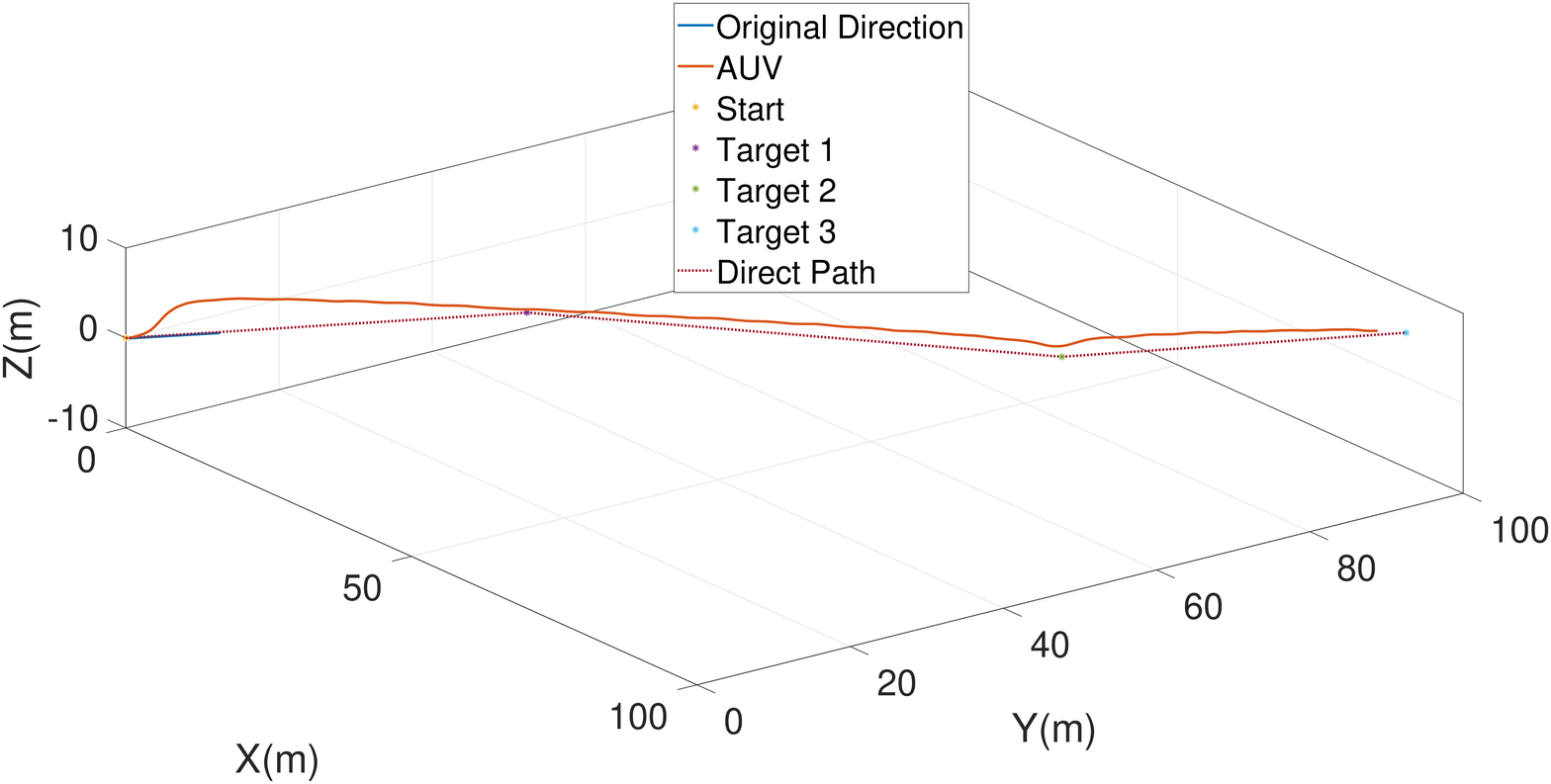}
\centering\caption{3D Waypoint tracking and depth control PILCO results with a target outside of the control depth.}
\centering
\label{DepthControl3D3points}
\end{figure*}

\cref{DepthControl3D} shows the results of policy 22. The policy is capable of keeping the depth at near to 5 meters and at the same time keep all four errors  at near zero value. Usually, the task described here will be done with two controllers a depth controller and a azimuth controller. As we are applying a single controller with two targets(elevation, depth) that at the start go on different paths. The $[X,Y,Z]$ results from the simulation are presented in \cref{DepthControlXYZ}. The policy produces a different behaviour than normal controllers, if by operator error a target was setup with a different depth than the desire depth the learned policy will try to push the vehicle to the target disregarding the depth by a small quantity such the target can be completed this can be observed in \cref{DepthControl3D3points} . Three targets were setups as $[30,30,5]$, $[70,70,3]$, and $[90,100,5]$ the second target is outside of the desire depth of the policy.


\subsection{Path Tracking}
\begin{figure*}[htbp]
\centering
\includegraphics[width=0.8\textwidth]{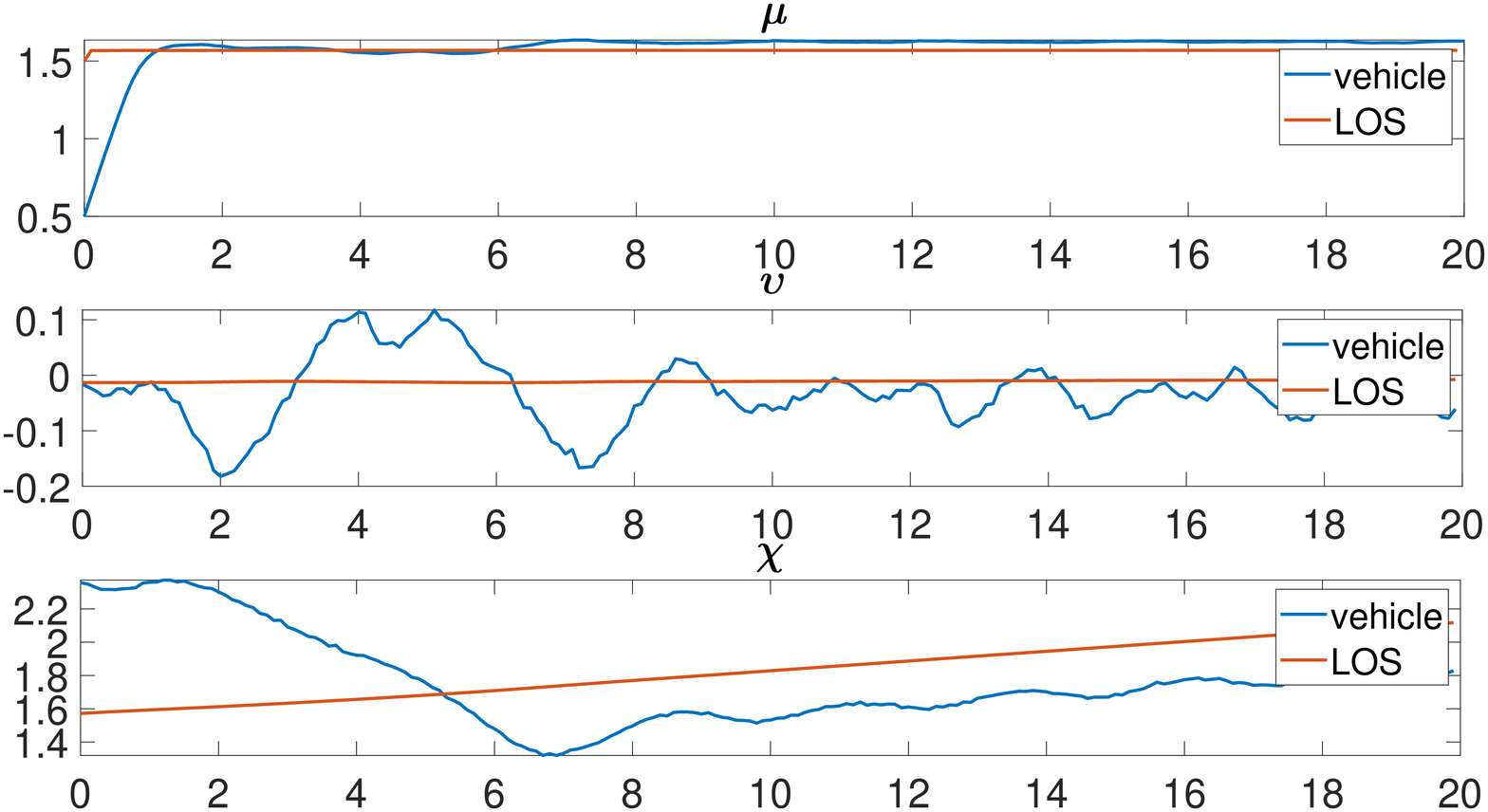}
\centering\caption{Sample of data used for learning of GPs model employed.}
\centering
\label{LOSPILCOSAMPLE}
\end{figure*}

\begin{figure*}
\centering
\includegraphics[width=0.8\textwidth]{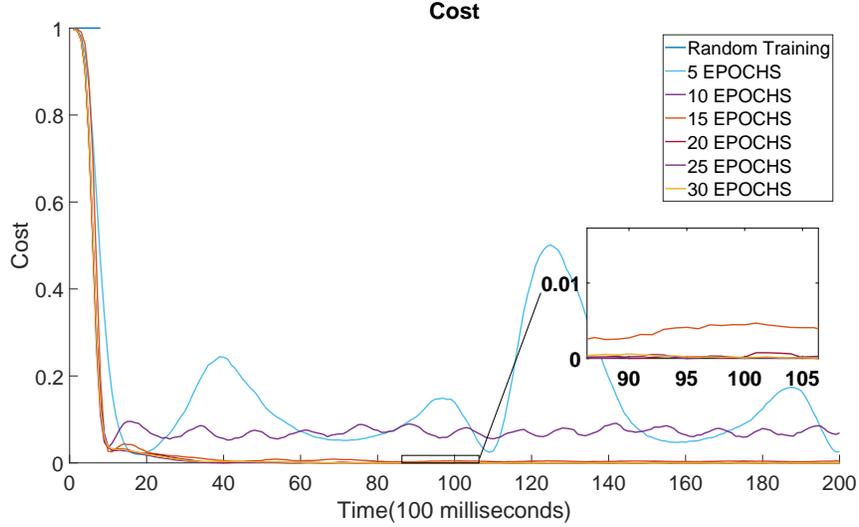}
\centering\caption{Cost evolution for each policy test over real model.}
\centering
\label{EpochsLOSPILCO}
\end{figure*}

\begin{figure*}[htbp]
\centering
\includegraphics[width=0.8\textwidth]{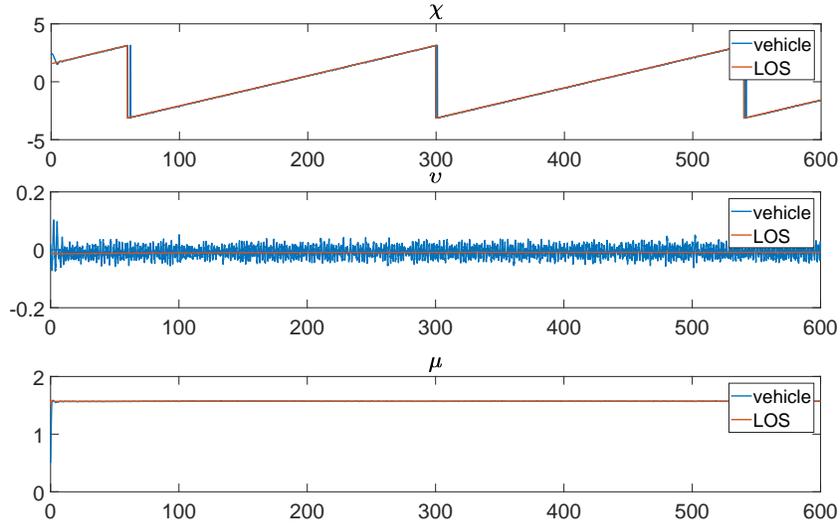}
\centering\caption{Desire LOS angles and speed and vehicle LOS angles and speed produced by PID controller.}
\centering
\label{PIDLOS}
\end{figure*}

\begin{figure*}[htbp]
\centering
\includegraphics[width=0.8\textwidth]{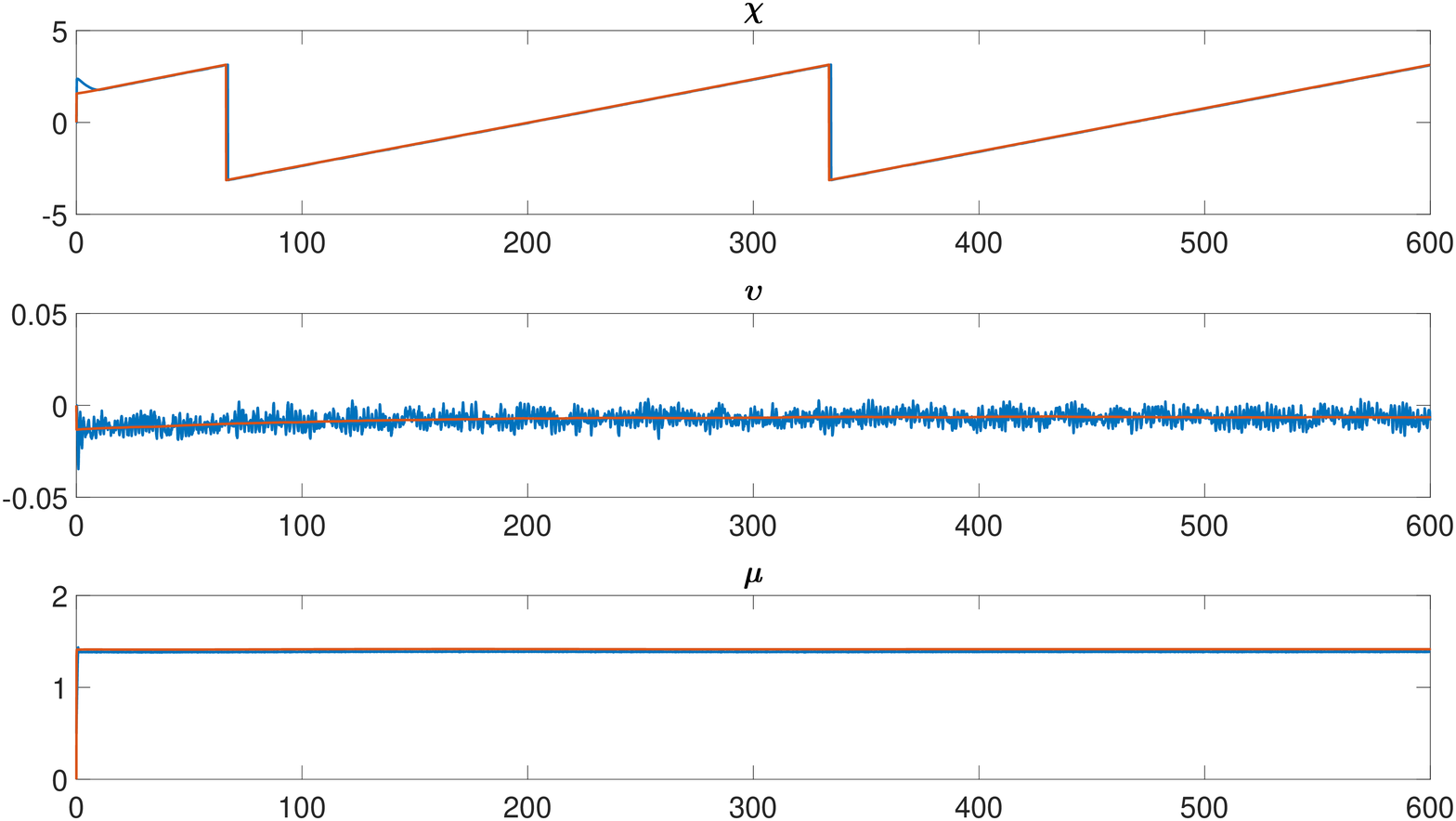}
\centering\caption{Desire LOS angles and speed and vehicle LOS angles and speed produced by PILCO policy.}
\centering
\label{PILCOLOS}
\end{figure*}

The training  evolution of the GPs model in the direction of a better policy allows PILCO to search for a better policy after each iteration with the real model (\cref{EpochsLOSPILCO}). \cref{LOSPILCOSAMPLE} presents an example of the data used for training of the GPs model that is later used for policy search.  The reinforced learning simulation to learn a LOS controller with PILCO shows that the reinforced learning algorithm can learn to follow the desired path based on the desire angles produced by a LOS law. The policy can perform better than a LOS-PID controller that is without noise compensation algorithms. The results from the PID to follow the LOS law can be seen in \cref{PIDLOS}, without noise the PID can stabilize himself very fast but with minimum noise in the depth sensor the $\chi$ angle cannot be completely stable and the controller will fluctuate. In the same way, PILCO controller (\cref{PILCOLOS}) reacts to the noise of the measurement but can follow the desired path in a better way that the LOS-PID controller implemented.

The measurement of the RMSE between the desire vehicle position and the vehicle position for both LOS-PID and LOS-PILCO is presented in \cref{LOSRMSE}. LOS-PILCO is shown to out-perform the PID controller in the reduction of error between the LOS law and the vehicle position. Not only in the accuracy of placing the vehicle in the correct position but in the speed of deployment of a controller to follow LOS will LOS-PILCO will require less tuning with field tests. Another advantage of PILCO is the absence of knowledge from the vehicle coefficients and not need to use other vehicles variables a surge, heave and sway speeds.

\begin{figure*}[htbp]
\centering
\includegraphics[width=0.8\textwidth]{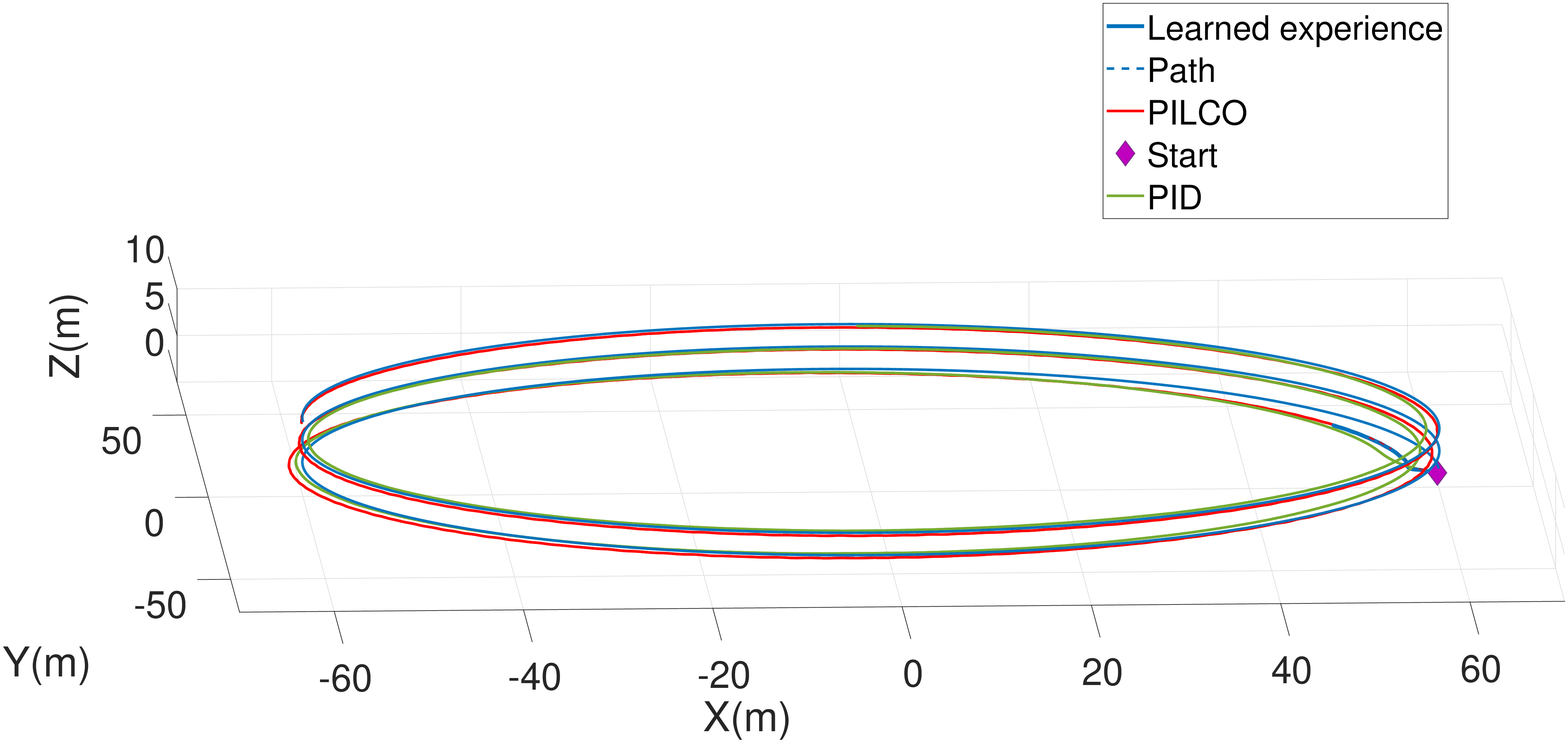}
\centering 
\caption{3D comparison of vehicle controlled with PID and PILCO controller.}
\centering
\label{LOSPILCORESULT3D}
\end{figure*}

\begin{figure*}[htbp]
\centering
\includegraphics[width=0.8\textwidth]{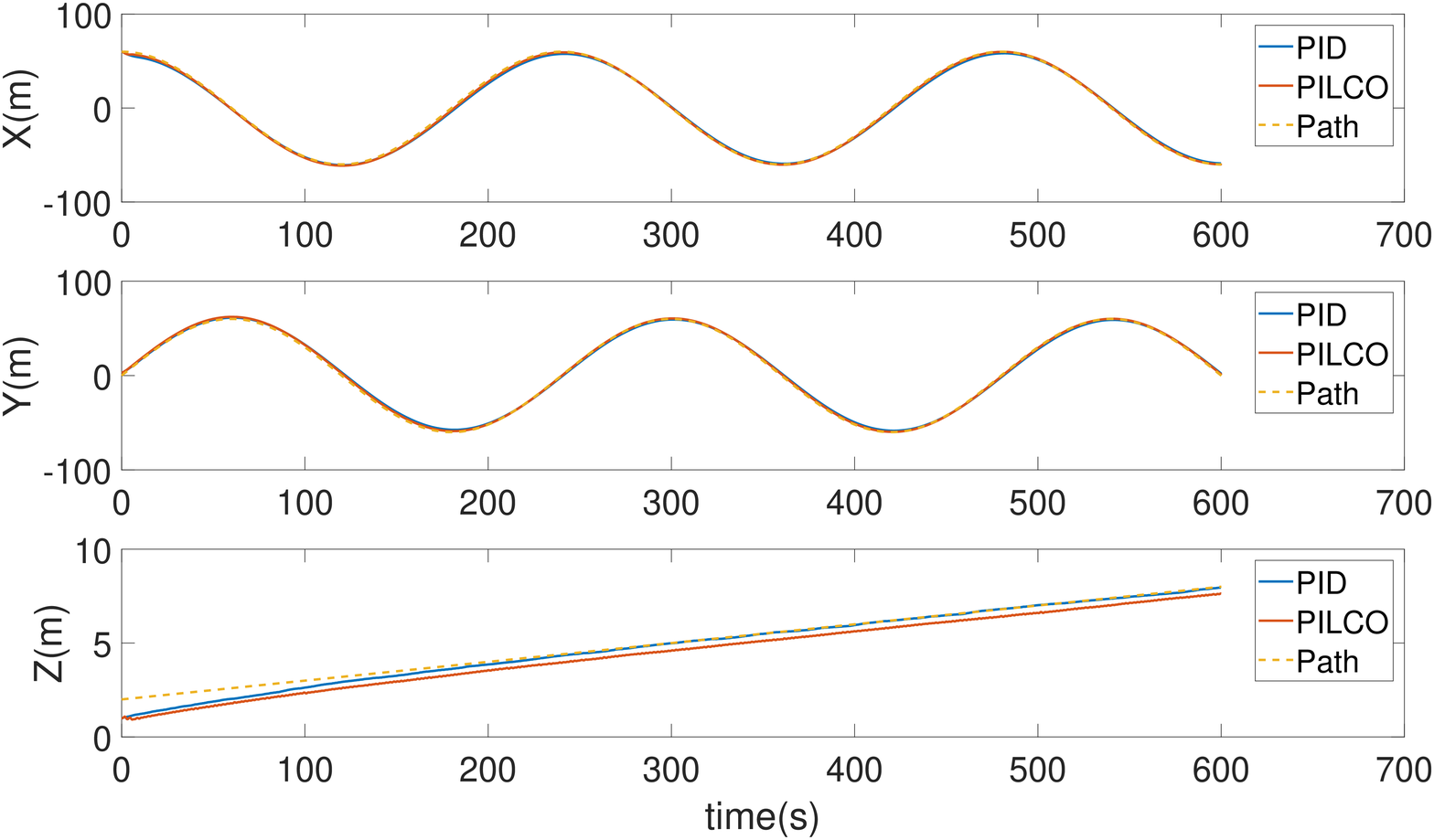}
\centering 
\caption{X,Y,Z comparison of vehicle controlled with PID and PILCO policy.}
\centering
\label{XYZPIDPILCO}
\end{figure*}

\begin{table}[htbp]
  \centering
  \caption{RMSE results from LOS-PID and LOS-PILCO.}
    \begin{tabular}{c|c}
    \textbf{RMSE} & \textbf{Value} \\
    \midrule
    LOS-PID & 1.214 \\
    LOS-PILCO & 0.89 \\
    \end{tabular}%
  \label{LOSRMSE}%
\end{table}%
\cref{LOSPILCORESULT3D} and \cref{XYZPIDPILCO} present the comparative plots of position in $X,Y,Z$ and the 3D path taken by the vehicle with LOS-PID controller and LOS-PILCO policy. Both controllers shows their ability to direct the AUV to the desired path, with the PILCO policy showing a better performance after episode 20.

\section{Conclusions}
This paper investigates the applicability of PILCO algorithms  and its requirements to learn policies to control underactuated AUVs. Three sets of simulations were designed to evaluate the capability of PILCO for waypoint tracking, depth control and path tracking. The simulations shown that a simple waypoint tracking control can be learnt in a small quantity of experiments over the real vehicle, the performance of the learnt policy was compared with a PID controller which is over-performed by the policy as the policy obtained is learned over the platform and the non-parametric model includes noise. 

In a similar way, a depth control policy was learned by mixing the waypoint tracking objective with the depth objective. However, in this research the learning of a policy to only control depth was not successful as the GPs model to allow the learning of a policy requires more information. Nevertheless, in the simulation of depth control it had been shown that a policy of depth control can be obtained by the learning of simultaneous waypoint tracking and depth control. The combination of objectives gives us a more intuitive behaviour, similar to what a human will do with the proposed objectives.

In the case of the proposed LOS-PILCO methodology, it has been shown that PILCO is a viable option to learn a policy to minimize the error between a LOS law and the vehicle position.  The PILCO policy is shown to perform equally and sometimes better than a PID. The RMSE shows that PILCO can obtain better performance and the long period of simulation shows that the learned policy can constantly minimize the error to the desired value.

PILCO algorithm has shown that is applicable to an underactuated AUV. In the simulations were consider limits that are a requirement for safety of the vehicle as maximum depth and maximum angles. The limits imposed to the platform do not limited the capability of PILCO to learn a viable policy. In the design of the cost function ,the vehicle shape and type of actuators force the selection of values of the cost function such that forward speed has to represent a higher cost than the specific vehicle angles or position. Model based reinforced learning for underactuated AUV shows to be a solution motion control of underwater vehicles and can be a solution to control bioinspired vechicles which are more dynamically complex to describe with a mathematical model.


\bibliographystyle{apsrev4-1}  

\bibliography{literature}

\begin{thebibliography}{10}

\bibitem{doi:10.5772/61037}
Zaopeng Dong, Lei Wan, Yueming Li, Tao Liu, Jiayuan Zhuang, and Guocheng Zhang.
\newblock Point stabilization for an underactuated auv in the presence of ocean
  currents.
\newblock {\em International Journal of Advanced Robotic Systems}, 12(7):100,
  2015.

\bibitem{980898}
F.~{Alonge}, F.~{D'Ippolito}, and F.~M. {Raimondi}.
\newblock Trajectory tracking of underactuated underwater vehicles.
\newblock In {\em Proceedings of the 40th IEEE Conference on Decision and
  Control (Cat. No.01CH37228)}, volume~5, pages 4421--4426 vol.5, Dec 2001.

\bibitem{RN1}
M.~Breivik and T.~I. Fossen.
\newblock Guidance-based path following for autonomous underwater vehicles.
\newblock {\em Oceans 2005, Vols 1-3}, pages 2807--2814, 2005.

\bibitem{6094949}
X.~Xiang, L.~Lapierre, C.~Liu, and B.~Jouvencel.
\newblock Path tracking: Combined path following and trajectory tracking for
  autonomous underwater vehicles.
\newblock In {\em 2011 IEEE/RSJ International Conference on Intelligent Robots
  and Systems}, pages 3558--3563, Sept 2011.

\bibitem{RN6}
Ahmad Forouzantabar, Babak Gholami, and Mohammad Azadi.
\newblock Adaptive neural network control of autonomous underwater vehicles.
\newblock {\em World Academy of Science, Engineering and Technology},
  6(7):304--309, 2012.

\bibitem{RN24}
K.~D. Do, J.~Pan, and Z.~P. Jiang.
\newblock Robust and adaptive path following for underactuated autonomous
  underwater vehicles.
\newblock {\em Ocean Engineering}, 31(16):1967--1997, 2004.

\bibitem{XIANG201514}
Xianbo Xiang, Lionel Lapierre, and Bruno Jouvencel.
\newblock Smooth transition of auv motion control: From fully-actuated to
  under-actuated configuration.
\newblock {\em Robotics and Autonomous Systems}, 67:14 -- 22, 2015.
\newblock Advances in Autonomous Underwater Robotics.

\bibitem{roper}
Chris Roper.
\newblock Using the gavia auv system to locate and document munitions dumped at
  sea.
\newblock Online, oct 2007.

\bibitem{holsen2015dune}
Sigurd~Andreas Holsen.
\newblock Dune: Unified navigation environment for the remus 100
  auv-implementation, simulator development, and field experiments.
\newblock Master's thesis, NTNU, 2015.

\bibitem{7587396}
R.~{Rout} and B.~{Subudhi}.
\newblock Narmax self-tuning controller for line-of-sight-based waypoint
  tracking for an autonomous underwater vehicle.
\newblock {\em IEEE Transactions on Control Systems Technology},
  25(4):1529--1536, July 2017.

\bibitem{ataei2015three}
Mansour Ataei and Aghil Yousefi-Koma.
\newblock Three-dimensional optimal path planning for waypoint guidance of an
  autonomous underwater vehicle.
\newblock {\em Robotics and Autonomous Systems}, 67:23--32, 2015.

\bibitem{saravanakumar2011waypoint}
S~Saravanakumar and T~Asokan.
\newblock Waypoint guidance based planar path following and obstacle avoidance
  of autonomous underwater vehicle.
\newblock In {\em ICINCO (2)}, pages 191--198, 2011.

\bibitem{RN29}
C.~Yu, X.~Xiang, and J.~Dai.
\newblock 3d path following for under-actuated auv via nonlinear fuzzy
  controller.
\newblock In {\em OCEANS 2016 - Shanghai}, pages 1--7, 2016.

\bibitem{XIANG2017165}
Xianbo Xiang, Caoyang Yu, and Qin Zhang.
\newblock Robust fuzzy 3d path following for autonomous underwater vehicle
  subject to uncertainties.
\newblock {\em Computers and Operations Research}, 84:165 -- 177, 2017.

\bibitem{7890302}
S.~Wang, Y.~Shen, Q.~Sha, G.~Li, J.~Jiang, J.~Wan, T.~Yan, and B.~He.
\newblock Nonlinear path following of autonomous underwater vehicle considering
  uncertainty.
\newblock In {\em 2017 IEEE Underwater Technology (UT)}, pages 1--4, Feb 2017.

\bibitem{RN12}
W.~Caharija, K.~Y. Pettersen, J.~T. Gravdahl, and E.~Borhaug.
\newblock Path following of underactuated autonomous underwater vehicles in the
  presence of ocean currents.
\newblock {\em 2012 Ieee 51st Annual Conference on Decision and Control (Cdc)},
  pages 528--535, 2012.

\bibitem{yang2016path}
Xiao Yang, Yue Shen, Kaihong Wang, Qixin Sha, Bo~He, and Tianhong Yan.
\newblock Path following for an autonomous underwater vehicle using gp-los.
\newblock In {\em OCEANS 2016-Shanghai}, pages 1--5. IEEE, 2016.

\bibitem{REPOULIAS20071650}
Filoktimon Repoulias and Evangelos Papadopoulos.
\newblock Planar trajectory planning and tracking control design for
  underactuated auvs.
\newblock {\em Ocean Engineering}, 34(11):1650 -- 1667, 2007.

\bibitem{liang2015path}
Xiao Liang, Yuan You, LF~Su, Wei Li, and Jundong Zhang.
\newblock Path following control for underactuated auv based on feedback gain
  backstepping.
\newblock {\em Technical Gazette}, 22(4):829--835, 2015.

\bibitem{gao2010global}
Jian Gao, Weisheng Yan, Ningning Zhao, and Demin Xu.
\newblock Global path following control for unmanned underwater vehicles.
\newblock In {\em Control Conference (CCC), 2010 29th Chinese}, pages
  3188--3192. IEEE, 2010.

\bibitem{doi:10.5772/64065}
Xiao Liang, Lei Wan, James~I.R. Blake, R.~Ajit Shenoi, and Nicholas Townsend.
\newblock Path following of an underactuated auv based on fuzzy backstepping
  sliding mode control.
\newblock {\em International Journal of Advanced Robotic Systems}, 13(3):122,
  2016.

\bibitem{RN28}
Z.~Chu and D.~Zhu.
\newblock 3d path-following control for autonomous underwater vehicle based on
  adaptive backstepping sliding mode.
\newblock In {\em Information and Automation, 2015 IEEE International
  Conference on}, pages 1143--1147, 2015.

\bibitem{6196982}
Xinqian Bian, Jiajia Zhou, Zheping Yan, and Heming Jia.
\newblock Adaptive neural network control system of path following for auvs.
\newblock In {\em 2012 Proceedings of IEEE Southeastcon}, pages 1--5, March
  2012.

\bibitem{8543571}
J.~Wang, C.~Wang, Y.~Wei, and C.~Zhang.
\newblock Three-dimensional path following of an underactuated auv based on
  neuro-adaptive command filtered backstepping control.
\newblock {\em IEEE Access}, pages 1--1, 2018.

\bibitem{1544973}
H.~Kawano.
\newblock Method for applying reinforcement learning to motion planning and
  control of under-actuated underwater vehicle in unknown non-uniform sea flow.
\newblock In {\em 2005 IEEE/RSJ International Conference on Intelligent Robots
  and Systems}, pages 996--1002, Aug 2005.

\bibitem{7483431}
H.~Shen and C.~Guo.
\newblock Path-following control of underactuated ships using actor-critic
  reinforcement learning with mlp neural networks.
\newblock In {\em 2016 Sixth International Conference on Information Science
  and Technology (ICIST)}, pages 317--321, May 2016.

\bibitem{yu2017deep}
Runsheng Yu, Zhenyu Shi, Chaoxing Huang, Tenglong Li, and Qiongxiong Ma.
\newblock Deep reinforcement learning based optimal trajectory tracking control
  of autonomous underwater vehicle.
\newblock In {\em Control Conference (CCC), 2017 36th Chinese}, pages
  4958--4965. IEEE, 2017.

\bibitem{fjerdingen2010auv}
Sigurd~A Fjerdingen, Erik Kyrkjeb{\o}, and Aksel~A Transeth.
\newblock Auv pipeline following using reinforcement learning.
\newblock In {\em Robotics (ISR), 2010 41st International Symposium on and 2010
  6th German Conference on Robotics (ROBOTIK)}, pages 1--8. VDE, 2010.

\bibitem{watkins1989learning}
Christopher John Cornish~Hellaby Watkins.
\newblock {\em Learning from delayed rewards}.
\newblock PhD thesis, King's College, Cambridge, 1989.

\bibitem{gaskett1999reinforcement}
Chris Gaskett, David Wettergreen, Alexander Zelinsky, et~al.
\newblock Reinforcement learning applied to the control of an autonomous
  underwater vehicle.
\newblock In {\em Proceedings of the Australian Conference on Robotics and
  Automation (AuCRA99)}, 1999.

\bibitem{6654139}
M.~P. Deisenroth, D.~Fox, and C.~E. Rasmussen.
\newblock Gaussian processes for data-efficient learning in robotics and
  control.
\newblock {\em IEEE Transactions on Pattern Analysis and Machine Intelligence},
  37(2):408--423, Feb 2015.

\bibitem{7401861}
M.~{De Paula} and G.~G. {Acosta}.
\newblock Trajectory tracking algorithm for autonomous vehicles using adaptive
  reinforcement learning.
\newblock In {\em OCEANS 2015 - MTS/IEEE Washington}, pages 1--8, Oct 2015.

\bibitem{RN76}
Thor~I Fossen.
\newblock {\em Guidance and control of ocean vehicles}, volume 199.
\newblock Wiley New York, 1994.

\bibitem{RN178}
Timothy Prestero.
\newblock {\em Verification of a six-degree of freedom simulation model for the
  REMUS autonomous underwater vehicle}.
\newblock Thesis, 2001.

\bibitem{gertler1967standard}
Morton Gertler and Grant~R Hagen.
\newblock Standard equations of motion for submarine simulation.
\newblock Technical report, DAVID W TAYLOR NAVAL SHIP RESEARCH AND DEVELOPMENT
  CENTER BETHESDA MD, 1967.

\bibitem{kim2006accurate}
Jinhyun Kim and Wan~Kyun Chung.
\newblock Accurate and practical thruster modeling for underwater vehicles.
\newblock {\em Ocean Engineering}, 33(5-6):566--586, 2006.

\bibitem{Fossen09}
Morten Breivik and Thor~I. Fossen.
\newblock Guidance laws for autonomous underwater vehicles.
\newblock In Alexander~V. Inzartsev, editor, {\em Underwater Vehicles},
  chapter~4. IntechOpen, Rijeka, 2009.

\bibitem{deisenroth2011pilco}
Marc Deisenroth and Carl~E Rasmussen.
\newblock Pilco: A model-based and data-efficient approach to policy search.
\newblock In {\em Proceedings of the 28th International Conference on machine
  learning (ICML-11)}, pages 465--472, 2011.

\bibitem{duvenaud2014automatic}
David Duvenaud.
\newblock {\em Automatic model construction with Gaussian processes}.
\newblock PhD thesis, University of Cambridge, 2014.

\bibitem{rasmussen2004gaussian}
Carl~Edward Rasmussen.
\newblock Gaussian processes in machine learning.
\newblock In {\em Advanced lectures on machine learning}, pages 63--71.
  Springer, 2004.

\bibitem{doi:10.1080/01691864.2014.888373}
Shusheng Bi, Chuanmeng Niu, Yueri Cai, Lige Zhang, and Houxiang Zhang.
\newblock A waypoint-tracking controller for a bionic autonomous underwater
  vehicle with two pectoral fins.
\newblock {\em Advanced Robotics}, 28(10):673--681, 2014.

\end{thebibliography}

\end{document}